\documentclass[aps,prx,twocolumn,floats,showpacs,superscriptaddress,nofootinbib]{revtex4-1}
\usepackage{graphicx,epsfig}
\usepackage{times,bbm}
\usepackage{graphics,dcolumn,bm,float}
\usepackage{amssymb,amsmath,rotate,color}
\usepackage[title,titletoc,toc]{appendix}
\usepackage{mathtools}
\usepackage{booktabs}
\usepackage{tcolorbox}
\usepackage[pagebackref=false,colorlinks,linkcolor=magenta,citecolor=blue,urlcolor=magenta]{hyperref}

\usepackage{wrapfig}
\usepackage{lipsum}
\usepackage{mwe}
\usepackage[mathlines]{lineno}
\usepackage{hyperref}
\usepackage{breqn}
\usepackage{bbold}
\usepackage{pgfplots}
\usepackage{float}
\usepackage{tkz-euclide}
\usepackage{braket}
\usepackage{physics}
\usepackage[caption=false]{subfig}
\usepackage[export]{adjustbox}
\usepackage{tikz}
\usetikzlibrary{through,calc}
\usetikzlibrary{positioning}

\newcommand{\be}{\begin{equation}}
\newcommand{\ee}{\end{equation}}

\newcommand{\bearr}{\begin{eqnarray}}
\newcommand{\eearr}{\end{eqnarray}}
\newcommand{\nn}{\nonumber}

\newcommand{\eps}{\varepsilon}

\newcommand{\bk}{{\boldsymbol{k}}}

\newcommand{\bv}{{\boldsymbol{v}}}
\newcommand{\br}{{\boldsymbol{r}}}

\newcommand{\bs}{\boldsymbol}
\newcommand{\bmt}{\left[\begin{matrix}}
\newcommand{\emt}{\end{matrix}\right]}

\begin{document}
\preprint{}
\title{Electric field assisted amplification of magnetic fields in tilted Dirac cone systems}

\author{S.A. Jafari}
\email{jafari@physics.sharif.edu}
\affiliation{
Department of Physics$,$ Sharif University of  Technology$,$ Tehran 11155-9161$,$ Iran
}

\date{\today}

\begin{abstract}
We show that the continuum limit of the tilted Dirac cone in materials such as $8Pmmn$-borophene and
layered organic conductor $\alpha$-(BEDT-TTF)$_2$I$_3$ deformation of the Minkowski spacetime of Dirac materials. 
From its Killing vectors we construct an emergent tilted-Lorentz (t-Lorentz) symmetry group for such systems. 
With t-Lorentz transformations we are able to obtain the exact solution of the Landau bands for a crossed configuration of electric and 
magnetic fields. For any given tilt parameter $0\le\zeta<1$ if the ratio $\chi=v_FB_z/cE_y$ of the crossed magnetic and electric fields that satisfies
$\chi\ge 1+\zeta$ one can always find appropriate t-boosts in both valleys labeled by $\pm$ in such a way the electric field can be t-boosted away, 
whereby the resulting pure effective magnetic field $B^\pm_z$ governs the Landau level spectrum around each valley. 
The effective magnetic field in one of the valleys is always larger than the
applied perpendicular magnetic field. This amplification comes at the expense of of diminishing the effective field in the opposite valley
and can be detected in various quantum oscillation phenomena in tilted Dirac cone systems. 
Tuning the ratio of electric and magnetic fields to $\chi_{\rm min}=1+\zeta$ leads to valley selective collapse of Landau levels.
Our geometric description of the tilt in Dirac systems reveals an important connection between the tilt and an incipient "rotating source" when
the tilt parameter can be made to depend on spacetime in certain way. 
\end{abstract}

\pacs{}

\keywords{}

\maketitle
\narrowtext

\section{Introduction}
Lorentz transformations that translate observations in two reference frames in such a way that $ds^2=-c^2dt^2+dx^2$ remains invariant~\cite{Arfken},
is at the heart of relativistic quantum field theories. The Lorentz symmetry (group) is part of the Poincar\'e group which is a fundamental symmetry of a world in which the 
elementary particles live. In this world, 
the speed $c$ of light is the universal upper limit of speeds, and is furthermore isotropic, giving rise to upright light cones. 
Now imagine an alternative world in which the light cones are tilted. This world does not exist in the standard model of particle physics,
but certain lattices solid state systems afford to mimic such spacetimes. 
In such a new world, the emergent symmetry at long wave-lengths is not the Lorentz symmetry, anymore. 
However such a spacetime is invariant under some deformation of the Lorentz group
which in this paper will be called the t-Lorentz. This new emergent symmetry group is parameterize by t-boosts where "t" 
emphasizes the importance of tilt in this strange world. 

Which condensed matter systems can realize such a continuous deformation of the Minkowski spacetime?
Owing to rich lattice structures of the solid state systems, the Dirac or Weyl equations~\cite{WehlingReview}
emerge as effective description of low-energy electronic degrees of freedom in one~\cite{Fradkin}, two~\cite{Novoselov2004} and 
three spatial dimensions~\cite{WehlingReview,Armitage2018,Fuseya2015}. 
The structure of the spacetime felt by electrons in a generic Dirac/Weyl material is precisely the Minkowski spacetime,
albeit the difference being that the speed of light will be replaced by Fermi velocity $v_F$ of the material at hand
which is usually 2-3 orders of magnitude smaller than $c$~\cite{Katsnelsonbook,Zabolotskiy2016,Suzumura2014}. 
The small ratio of $v_F/c$ does not harm the Lorentz symmetry\footnote{It has more interesting effect of enhancing the fine structure constant
for Dirac fermions of condensed matter~\cite{Katsnelsonbook}.}. The life becomes more interesting when it comes to 
lattices with non-symmorphic symmetry elements. After all, the condensed matter systems are mounted on a 
background crystal, not on the vacuum. Therefore, the Lorentz symmetry is not necessarily the symmetry group of
linear band touching in solid state systems. In fact, 
the rich point group symmetry of crystals can provide classes of fermions which do not necessarily have any counterpart 
in particle-physics~\cite{BernevigNexus}. For example unconventional fermions living on crystals with non-symmorphic 
symmetry elements can boldly violate the spin-statistic theorem~\cite{Peskin} which rests on the Lorentz symmetry~\cite{Pauli1940}. 
Therefore, relaxing the Lorentz symmetry seems to produce opportunities not availalbe in physics of elementary particles. 

In this paper we would like to show that 
the effect of non-symmorphic lattice structures is not limited to generation of strange forms of fermions in condensed matter. 
When the Dirac/Weyl equations are brought to their mundane sub-eV solid-state framework, 
the non-symmorphic symmetry elements which arise from the underlying lattice can generate 
finite amount of tilt in the {\it intrinsic} Dirac cone spectrum of electronic degrees of freedom~\cite{TohidBorophene}. 
Such a finite tilt should be contrasted to very small tilt that can be {\it extrinsically} induced in
graphene by appropriate strains~\cite{Cabra2013,Mao}. This solid-state based world of tilted Dirac cones is the subject of the present paper. 
The tilt deformation of Dirac equation although destroys the Lorentz symmetry, but still a deformed
version of Lorentz group survives in the form of t-Lorentz group. The purpose of this work is: (1) to 
study the isometries of the spacetime with tilted Dirac cones and construct the t-Lorentz group and identify its
algebraic structure and (2) to show that the tilt can help to {\em amplify} the effective magnetic field in
one of the valleys. This is achieved by exactly solving the Landau band problem in crossed magnetic and electric
field background which can only be neatly done by t-Lorentz transformations and have no analog in tilt-less Dirac
systems. 

The candidate materials related to the spacetime discussed in our work are quasi-two dimensional (molecular orbital based) 
systems such as organic $\alpha$-(BEDT-TTF)$_2$I$_3$~\cite{Suzumura2006,Tajima2006}, or an atom thick sheet of
$8Pmmn$ borophene~\cite{TohidBorophene}. 
The advantage of the later system besides being in two space dimensions which offers functionalization and manipulation
opportunities is that: (i) Its intrinsic tilt parameter is quite large. (ii) The tilt can be further
controlled with perpendicular electric field from the under-tilted regime of the pristine borophene to over-tilted regime 
(iii) The particular non-symmorphic structure of the space group protects the Dirac node~\cite{Hatsugai} 
as long as the spin-orbit coupling is small. Owing to very small atomic number of Boron, 
the intrinsic spin-orbit interaction is $\sim 0.02$~\cite{Yao-Cat} meV. So for all practical purposes, 
the intrinsic tilted Dirac cone in pristine borophene can be assumed to be massless. 
An essential feature of tilted Dirac cones is that being mounted on a lattice, they always come in pairs
with opposite tilts, $\pm \zeta$. This sign difference is behind the amplification mechanism that we will discuss in this paper. 

\section{The t-Lorentz transformations}
Let us start by minimal form of tilted Dirac equations for one of the valleys~\cite{Goerbig2008,Tohyama2009,SaharTilt1,SaharTilt2},
 \be
 H=\hbar v_F \begin{pmatrix}  \zeta k_x &  k_x-ik_y\\   k_x+ik_y &  \zeta k_x  \end{pmatrix}=\hbar v_F (\zeta k_x \tau_0+\bk.\bs{\tau}).
     \label{nmatrixform}
 \ee 
where the Pauli matrices $\tau_\mu$ with $\mu=0,1,2$ act on the orbital space and $\tau_0$ is the unit $2\times 2$ matrix in this space.
This theory is characterized by two velocity scales: $v_F$ determines the cone like dispersion, while $v_t=\zeta v_F$ determines the
tilt of the energy axis with respect to the $k_xk_y$ plane. We have used our freedom too choose coordinate system such that the $k_x$ axis is along the tilt direction.
The pristine borophene with intrinsic $\zeta\sim 0.4$ lies in the under-tilted regime (i.e. $0<\zeta<1$). 
It can be tuned by a perpendicular electric field to over-tilted regime with $1<\zeta$~\cite{TohidBorophene}.
From the effective theory of $8Pmmn$-borophene it follows that the other valley is obtained by $\zeta\to -\zeta$ and $\tau_x\to -\tau_x$. 
The anisotropy of the Fermi velocity $v_F$ in any realistic material~\cite{Katayama2008,Suzumura2014} can be removed by a 
rescaling of momenta (or coordinates) which will give rise to a constant Jacobian and does not alter the physics~\footnote{
When the Fermi velocity scale is random, interesting "gravitational lensing"-like phenomena appear in tilted $2+1$-dimensional
Dirac systems~\cite{Ghorashi}.}.
The eigenvalues and eigenstates of  the tilted Dirac cone Hamiltonian are given by,
\be
E_s(\bk)=  k (s+\eta \cos {\theta}_{\bk}) ~~~,~~~
 \ket{\bs{k},\pm}= \frac{1}{\sqrt{2}}\begin{pmatrix} 1 \\ \pm e^{i\theta_{\bk}} \end{pmatrix},
 \label{dispersion.eqn}
 \ee
where $s=\pm 1$ refers to positive ($E_+$) and negative ($E_-$) energy branches, and 
$\theta_{\bk} $ is polar angle of the wave vector, $\bk$, with respect to the $x$ direction. 

Following Volovik~\cite{Volovik2016,Volovik2018}, the dispersion of a tilted Dirac cone can be 
viewed as a null-surface in a Painelev\'e-Gullstrand spacetime,
\be
   ds^2=-v_F^2 dt^2+(d\br-\bv_t dt)^2.
   \label{PG.eqn}
\ee
The space part of this metric have acquired a Galilean boost by velocity $\bv_t$. 
The dispersion relation of massless particles in this spacetime is given by $g^{\mu\nu}k_\mu k_\nu=0$ which upon identifying
$k_\mu=(E/v_F,\bk)$ gives $(E-\bk.\bv_t)^2-v_F^2k^2=0$. This is nothing but the dispersion relation~\eqref{dispersion.eqn}
of the tilted Dirac fermions. {\it Therefore for the tilted Dirac fermions, the spacetime is given by metric~\eqref{PG.eqn}.}
When the tilt velocity $v_t=\zeta v_F$ is can be made to depend on the radial coordinate~\cite{TohidBorophene}, 
the condition $v_t>v_F$ for over-tilted Dirac cone corresponds to the black-hole in 3+1 dimensions, and BTZ black-holes~\cite{BTZ} in 2+1 dimensional spacetime geometry. 
The explicit form of this metric in 1+1 dimension is given by,
\be
   g_{\mu\nu}=\bmt
   -\lambda^2	&-\zeta \\
   -\zeta	&1
   \emt
   \leftrightarrow
   g^{\mu\nu}=\bmt
   -1		&\zeta\\
   \zeta	&\lambda^2
   \emt,
   \label{PGmetric2.eqn}
\ee
where we have introduced $\lambda^2=1-\zeta^2$. 
In this work we confine ourselves to a much simpler form of this metric where the tilt velocity $\bv_t$ is constant
all over the spacetime, and as such is a flat spacetime. So in this work we will study the fundamental
symmetry and physics arising from a constant tilt parameter $\zeta$. 
Further evidence in favour of the above geometry comes from the fact that the brute force
calculation of the polarization tensor $\Pi^{\mu\nu}\sim\langle j^\mu j^\nu\rangle$ shows that
it acquires the govariant form $(g^{\mu\nu}q^2-q^\mu q^\nu)\pi(q)$, {\em only} when the tilted
geometry is used~\cite{SaharCovariant}
From now on we will assume that the Fermi velocity $v_F=1$ and will restore it when required. 

For clarity, let us first derive the t-Lorentz transformation in 1+1 dimensions. 
Since we will need to ensure that $\zeta=0$ reduces to the standard Minkowski spacetime,
let us start by a quick reminder of the Lorentz transformation in this space: In this case a small Lorentz transformation 
parameterized by $\kappa$ is $\Lambda_0={\mathbbm 1}+i\kappa K_0$, where $K_0$ is the generator of transformation. 
Invariance of $ds^2=-dt^2+dx^2$ (equivalent to $g_0={\rm diag}(-1,1)$ metric) means $g_0 K_0+K_0^Tg_0=0$.
This fixes $iK_0=\tau_1$, where $\tau_1$ is the first Pauli matrix~\cite{Peskin,ZeeQFTBook}. From this the Lorentz 
transformation for finite $\kappa$ in 1+1 dimension becomes,
\be
   \Lambda_0(\kappa)=\bmt
      \cosh(\kappa)	&\sinh(\kappa)	\\
      \sinh(\kappa)	&\cosh(\kappa)	\\
   \emt
   \label{lorentz0.eqn}
\ee
where the boost parameter $\kappa$ is related to the velocity $\beta$ by 
$\cosh(\kappa)=(1-\beta^2)^{-1/2}\equiv \gamma$ and $\sinh(\kappa)=-\beta \gamma $. 
The same logic allows us to derive a generalized Lorentz transformation in $1+1$-dimensions
in presence of a non-zero tilt parameter $\zeta$. We expand $\Lambda={\mathbbm 1}+i\kappa K$, and
require it to leave the metric~\eqref{PGmetric2.eqn} of tilted Dirac fermions invariant.
Again this fixes the generator of the t-Lorentz transformation,
\be
   iK=\bmt
      -\zeta		&1\\
      \lambda^2		&\zeta
   \emt.
\ee
From this one can immediately find the large t-Lorentz transformation. For a t-boost along $x$ (tilt) direction we obtain,
\bearr
   \Lambda^x_{1+1}
   =&&\gamma\bmt
      1+\zeta\beta	&-\beta		\\
      -\lambda^2\beta	&1-\zeta\beta	\\
   \emt.
   \label{lorentzeta.eqn}
\eearr
Needless to say, for $\zeta=0$ this equation reduces to Eq.~\eqref{lorentz0.eqn}. 

Now we are ready to construct the t-Lorentz transformations in $2+1$ dimensions. The coordinates 
are defined by $x^\mu=(x^0=v_F t,x^1=x,x^2=y)$. We choose the $x$ direction along the tilt direction
such that the tilt is given by the two-vector ${\bs\zeta}=(\zeta_1=\zeta,\zeta_2=0)$. For this choice
the metric in $2+1$ dimensions will be a generalization of Eq.~\eqref{PGmetric2.eqn} and is given by~\cite{SaharPi},
\be
   g_{\mu\nu}=\bmt
   -\lambda^2	&-\zeta 	&0\\
   -\zeta	&1		&0\\
   0		&0		&1
   \emt
   \label{PGmetric3.eqn}
\ee

For a t-boost along the $x$ direction where the tilt lies, the corresponding $2+1$-dimensional generator 
is obtained from the $1+1$-dimensional generator by padding with $0$s, while the other boosts are 
obtained from $gK+K^Tg=0$ as,
\bearr
   iK_x=\bmt
      -\zeta		&1		&0\\
      \lambda^2		&\zeta		&0\\
      0			&0		&0
   \emt,~
   iK_y=\bmt
      0			&0		&1\\
      0			&0		&0\\
      \lambda^2		&\zeta		&0
   \emt,~
   iJ_z=\bmt
      0			&0		&0\\
      0			&0		&1\\
      \zeta		&-1		&0
   \emt\nn
\eearr
The names are chosen to emphasize that in the limit $\zeta=0$ they reduce to corresponding 
generators of boosts along, $x$, $y$ and rotation around $z$ axes, respectively. 
The algebra of these generators is a deformation of Lorentz algebra~\cite{Padmanabhan},
\bearr
   && [K_x,K_y]=i\zeta K_y-i(1-\zeta^2)J_z,\\
   && [J_z,K_x]=iK_y+i\zeta J_z,\\
   && [J_z,K_y]=-iK_x,
\eearr
and in the limit $\zeta=0$ reduces to what one expects for the Lorentz group~\cite{Ryder,Padmanabhan}. 

With the above generators we can construct the t-Lorentz transformations along $x$ (tilt direction) and $y$ (transverse to tilt direction)
by simply exponentiating $iK_x \kappa$ and $iK_y\kappa$. Similarly the t-rotation around $z$ axis will be obtained
by exponentiation of $iJ_z\theta$. The result is,
\bearr
   \Lambda^x_{2+1}
   =&&\bmt
      \cosh\kappa-\zeta\sinh\kappa	&\sinh\kappa			&0\\
      \lambda^2\sinh\kappa		&\cosh\kappa+\zeta\sinh\kappa	&0\\
      0					&0				&1
   \emt,
   \label{lorentzetax3.eqn}\\
   \Lambda^y_{2+1}
   =&&\bmt
      \cosh\lambda\kappa		&\frac{\zeta}{\lambda^2}\left[\cosh\lambda\kappa-1\right]	&\frac{1}{\lambda}\sinh\lambda\kappa\\
      0					&1								&0\\
      \lambda\sinh\lambda\kappa	&\frac{\zeta}{\lambda}\sinh\lambda\kappa			&\cosh\lambda\kappa
   \emt,
   \label{lorentzetay3.eqn}\\
   &&R(\theta)=
   \bmt
      1				&0		&0\\
      \zeta(1-\cos\theta)	&\cos\theta	&\sin\theta\\
      \zeta\sin\theta		&-\sin\theta	&\cos\theta
   \emt.
   \label{Rotationzetaz3.eqn}
\eearr
The first equation, namely Eq.~\eqref{lorentzetax3.eqn} is straightforward generalization of Eq.~\eqref{lorentzeta.eqn}. 
The second Eq.~\eqref{lorentzetay3.eqn} also reduces to the standard Lorentz transformation of the Minkowski spacetime in the limit of $\zeta=0$ ($\lambda=1$). 
The third Eq.~\eqref{Rotationzetaz3.eqn} is also a generalization of rotation around $z$ axis which again reduces to the
rotation in the Minkowski space in the limit of $\zeta=0$. 

In the appendix we give a more systematic derivation of the above transformations using the
method of Killing vectors to ensure that the above derivation does not miss any isometry of the spacetime of 
tilted Dirac fermions.

\section{Amplification of magnetic fields}
One of the fascinating results of the special theory of relativity is that the electric and
magnetic fields are actually components of the same tensor which covariantly transforms
under the Lorentz transformation. Lukose and coworkers have used this fact to obtain a
beautiful exact solution of the Landau bands in crossed $E_y$ and $B_z$ fields. 
In a tilted Minkowski space with metric~\eqref{PGmetric3.eqn} if the electric and magnetic fields are due to 
sources living in the same sapce, the field strength tensor $F^{\mu\nu}$ will also transform 
under t-Lorentz transformations. The nice property of two-dimensional Dirac materials is that
the Fermi surface can be tunned to regimes where hydrodynamic regime with emergent electromagnetic
fields can be achieved. The t-Lorentz transformations enables us to study the Landau bands in crossed electric and 
magnetic fields in $2+1$-dimensional tilted Dirac cone systems such as borophene. 

To study the transformation of emergent electric and magnetic fields under t-Lorentz transformation let us
restore the Fermi velocity $v_F$ to manifestly see its interplay with the speed of light, $c$. 
Assuming a $B$-field along $z$ direction and an $E$-filed along $y-$ direction in borophene, 
then a t-boost along the $x$-direction by the velocity $\beta v_F$ changes $(E_y,B_z)$ according to~\cite{Vinu2007,Katsnelsonbook},
\bearr
E'_y             &=&\gamma \left[(1+\zeta\beta)E_y-\beta\frac{v_F B_z}{c} \right],\\
\frac{v_F}{c}B'_z&=&\gamma \left[(1-\zeta\beta)\frac{v_F B_z}{c}-(1-\zeta^2)\beta E_y \right].
\eearr
Again the $\zeta=0$ limit agrees with corresponding result in graphene~\cite{Vinu2007}.
When the electromagnetic fields do not arise from electric charges in spacetime~\eqref{PGmetric3.eqn}
one has to set $\zeta=0$ in the transformation of $E$ and $B$ fields. 
In the limit of $\zeta=0$, a boost along $x$-direction does not change $E_x$~\cite{Ryder}. 
The same holds for a t-Lorentz transformation along the $x$ (tilt) direction. 
For a t-Lorentz transformation $\Lambda=\Lambda^x_{2+1}$ with arbitrary $\zeta$ we have:
\bearr
   {E'}_x&=&{F'}^{01}=\Lambda^0_\rho\Lambda^1_\sigma F^{\rho\sigma}=\Lambda^0_0\Lambda^1_1 F^{01}+\Lambda^0_1\Lambda^1_0 F^{10}\nn\\
       &=& \gamma^2(1+\beta\zeta)(1-\beta\zeta)E_x+(-\beta)(-\lambda^2\beta)(-E_x)=E_x,\nn
\eearr
which is similar to the standard Lorentz transformation. 

Following Lukose and coworkers~\cite{Vinu2007}, we choose the t-boosted frame such in the t-boosted frame 
the electric field can be eliminated. This can be achieved for the t-boost parameter,
\be
   \beta_{*\pm}=\frac{E_y}{v_FB_z/c\mp\zeta E_y}=\frac{1}{\chi\mp\zeta},~~~\chi=\frac{v_FB_z}{cE_y},
\ee
where we have restored the valley index $\pm$ and have used the fact that the sign of the tilt parameter $\zeta$
for the two valleys is opposite. For a given material the $\zeta$ is fixed (which can be assumed to be positive) 
and for type-I tilted Dirac systems is further less than $1$. In this case the condition $\chi>1\pm\zeta$ implies that $|\beta_{*,\pm}|<1$ and
hence a (separate) t-boost (for each valley with parameters $\beta_{*,\pm}$) can be found that eliminates the electric field. 
In such a frame a purely effective magnetic field of the following form will be felt,
\be
   B^{\pm}_z=B_z\frac{\sqrt{1-\beta_{*,\pm}^2}}{1\pm\zeta\beta_{*,\pm}}=B_z\frac{\sqrt{(\chi\mp\zeta)^2-1}}{\chi}.
\ee
The condition $\chi>1+\zeta$ guarantees that both t-boost parameters $\beta_{*,\pm}$ have magnitudes less than one
and pure magnetic fields $B_z^\pm$ in both valleys can be realized. In particular at $\chi_{\rm min}=1+\zeta$ the $B_{z}^+$ vanishes,
while $B_{z}^-$ becomes $2B_z\sqrt{\zeta/(1+\zeta)}$. The vanishing of $B_{z}^+$ at this particular value of $\chi_{\rm min}$ implies that the Landau orbits around the $+$ valley
collapse~\cite{Vinu2007} while the Landau orbits around the other valley survive. The magnetic field around "$-$" valley will be
larger than $B_z$ when $\zeta>1/3$. This is in contrast to the situation in graphene with $\zeta=0$, where the behavior of 
Landau levels in both valleys is the same. By tuning the ratio $\chi$ of crossed $B_z$ and $E_y$ fields beyond $\chi_{\rm min}$, 
the $B_z^+$ starts to increase from zero, but always remains less than $B_z^-$. By reversing the direction of either $E_y$ or $B_z$ 
which amounts to flipping the sign of $\chi$, the collapsed Landau levels will be centered around the other valley.

So far the Landau levels are calculated in the t-boosted frame where $E_y$ is zero. 
The Landau levels in the boosted frame are, $\eps^\pm_n=\lambda\hbar\omega_c\sqrt n$ where the cyclotron frequency is
$\hbar\omega^\pm_c=\sqrt{2\pi eB^\pm_z/\hbar c}$~\cite{Tohyama2009}. 
To obtain the Landau bands that are observed
in the laboratory frame, one must t-boost back the Landau energy-momentum $3$-vector $(\eps^\pm_n,0,0)$~\cite{Vinu2007}. 
These levels will acquire a dispersion in the laboratory frame as follows~\cite{Vinu2007}:
$(\eps'_{n},k'_x,k'_y)=\gamma_{*,\pm}\eps^\pm_n(1\mp\zeta\beta_{*,\pm},-\lambda^2,0)$ which implies 
\be
   \eps'_{n}=\frac{1\pm\zeta\beta_{*,\pm}}{(\zeta^2-1)\beta_{*,\pm}} k'_x
   =\frac{\chi}{\zeta^2-1}k'_x.
\ee
The velocity related to dispersion along $x$ (tilt) direction is controlled by the ratio $\chi$ and diverges for $\zeta\to 1$. 

\begin{figure}[t]
   \includegraphics[width =0.42\textwidth]{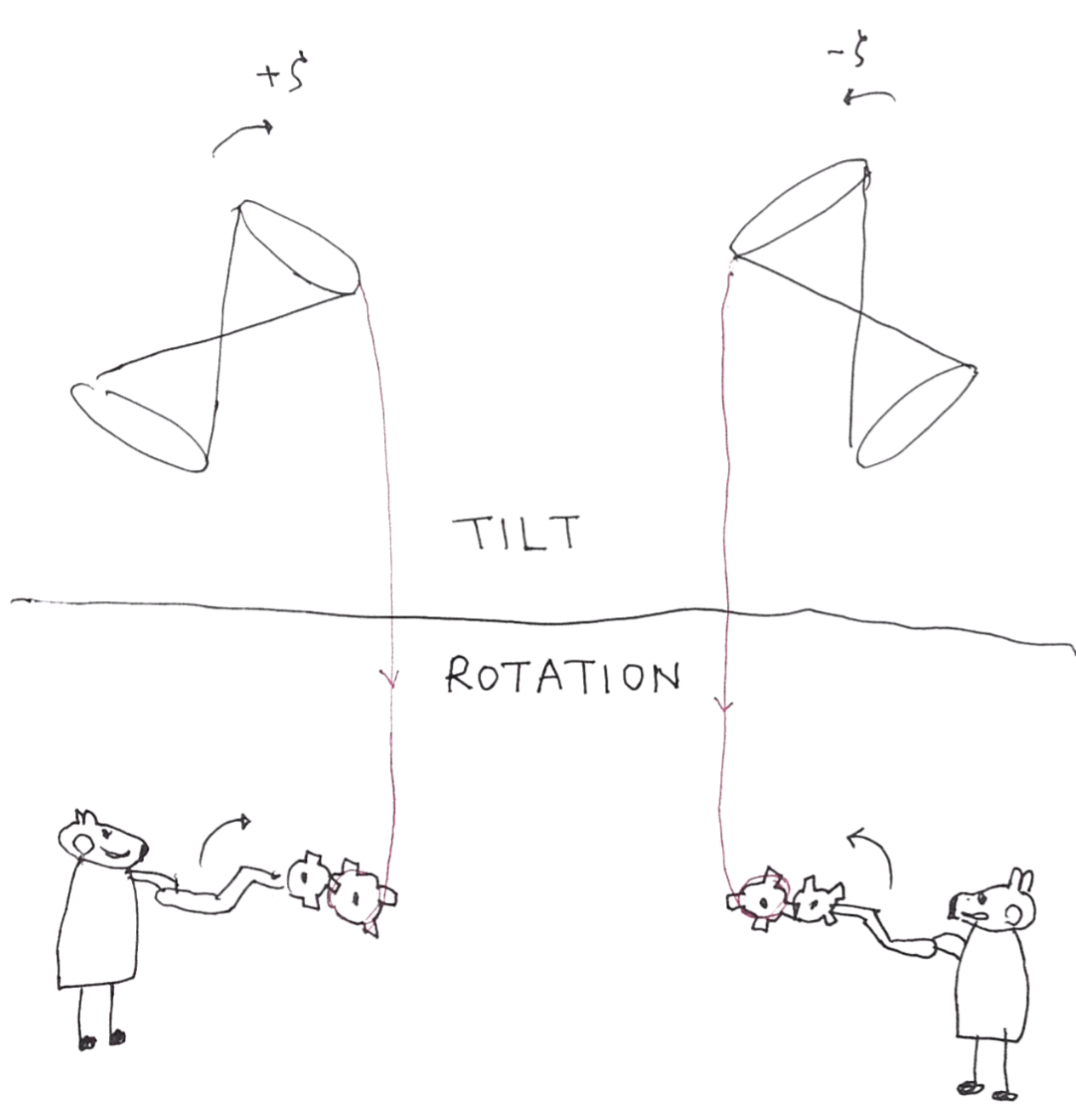}
   \caption{(Color online) Connection between tilt of the Dirac cone and rotating source in the metric. Sketch by Hasti Jafari.}
   \label{hasti.fig}
\end{figure}
The asymmetry between the Landau quantization of the two valleys is due to the tilt parameter $\zeta$. Indeed the metric~\eqref{PGmetric3.eqn}
in the small $\zeta$ limit where $\zeta^2$ effects can be ignored coincides with the metric of a rotating gravitational source if
the $\zeta$ can be made to depend in space according to $\zeta_x\propto y$ and $\zeta_y\propto -x$~\cite{Ryder}.
The effect of rotation of the source is such that it gives rise to non-zero $g_{0i}$ components in the metric~\cite{Ryder}.
These components are responsible for Lens-Thirring effect which is the precision of spins arising from the vector field ${\bs\zeta}$~\cite{Ryder}.
In this limit, for all practical purposes the role of ${\bs\zeta}$ will be formally equivalent to a "magnetic" field given by $\nabla\times{\bs\zeta}$~\cite{Ryder}.
The opposite ${\bs\zeta}$ for the two valleys creates opposite "gravitomagnetic" effects. For ${\bs\zeta}\propto (y,-x)$ it would not be
surprising that the tilt parameter $\zeta$ helps to diminish the effective magnetic field in one valley and
amplify it in the other valley. Given that in the present case $\bs\zeta$ does not depend on spacetime, the above amplification
of emergent magnetic fields remains a genuine property of t-Lorentz covariance. Note that
in the case of graphene ($\zeta=0$) the effective magnetic field is always less 
than the applied $B_z$, while in the present case the magnetic field felt in the t-boosted frame for one of the valleys can be enhanced. 


\section{Summary and discussion} Isometries of the metric compatible with tilted Dirac cone are
tilted-Lorentz transformation. Transformation of the emergent electromagnetic field strength tensor under t-Lorentz 
transformations allows us to exactly solve the problem of tilted Dirac fermions in crossed electric and
magnetic fields. The solution consists of two interpenetrating Landau bands. The Landau levels $\eps_n^\pm$ in the 
two valleys $\pm$ are asymmetric. The effective magnetic field felt in one valley is always smaller than the background $B_z$
while in the other valley stronger magnetic field is felt. This amplification has no analog in upright Dirac cone systems.
The way the tilt parameter appears in the metric is similar to the metric of a rotating gravitational source (See Fig.~\ref{hasti.fig}) 
if the tilt parameter depends on space coordinate as ${\bs\zeta}\propto (y,-x)$. This can be achieved in $8Pmmn$ borophene~\cite{TohidBorophene}. 
In this way the spatio-temporal variation~\cite{TohidBorophene} of the tilt vector $\bs\zeta$ is expected to couple to the spin of the electrons
via spin-rotation coupling~\cite{MaekawaSpinRotation}. This is expected to give rise to electric field control of spin currents that employ that
couple through the metric of the spacetime~\cite{MashhoonSpinEarth}. In the light of recent reports of torsional anomaly 
in Weyl semimetals with tilted cone~\cite{TorsionalAnomaly} and gravitational lensing-like effect in tilted Dirac cone~\cite{Ghorashi}, 
and our recent calculation of the covariant structure of the polarization tensor in tilted cone systems~\cite{SaharPi} it appears
that the tilted Dirac/Weyl cone systems are a fertile search grounds for gravitational analogies in condensed matter. 

\section{Acknowledgements} I thank S. Baghram and B. Mashhoon for allowing me to attend their inspiring courses on general relativity. 
Useful comments from B. Mashhoon, M. M. Sheikh-Jabbari and Mehdi Kargarian, Delaram Mirfendereski and Ali Mostafazadeh is appreciated. 
This work was supported by research deputy of Sharif University of Technology, grant no. G960214 and the
Iran Science Elites Federation. I thank Hasti Jafari for her assistance with preparation of Fig.~\ref{hasti.fig}.


\bibliography{mybib}

\newpage
\onecolumngrid

\appendix
\section{Killing vectors}
Killing vectors satisfy the Killing equation~\cite{Ryder},
\be
   g_{\mu\nu,\lambda}\xi^\lambda+g_{\mu\lambda}\xi^\lambda_{~,\nu}+g_{\lambda\nu}\xi^\lambda_{~,\mu}=0
   \label{Killing.eqn}
\ee
For simplicity let us first work out the Killing vectors for the $1+1$-dimensional space with 
a constant tilt parameter $\zeta$. The metric is given by Eq.~\eqref{PGmetric2.eqn}. Using this metric
in the Killing equation~\eqref{Killing.eqn}, for $\mu\nu=00$ will then give,
\be
   g_{0\lambda}\xi^\lambda_{~,0}=0 \Rightarrow (\zeta^2-1)\xi^0_{~,0}=\zeta\xi^1_{~,0}\nn
\ee
which implies
\be
   \xi^1=\frac{\zeta^2-1}{\zeta}\xi^0+b(x)
   \label{xi0.eqn}
\ee
with $b(x)$ some yet unknown function of $x$ only. Similarly the $\mu\nu=11$ case gives $g_{1\lambda}\xi^\lambda_{~,1}=0$
which implies $\zeta\xi^0_{~,1}=\xi^1_{~,1}$ from which it follows that,
\be
   \xi^1=\zeta\xi^0+c(t)
   \label{xi1.eqn}
\ee
with $c(t)$ a yet unknown function of $t$ only. Combining Eq.~\eqref{xi0.eqn} and~\eqref{xi1.eqn} gives,
\be
   \xi^0=\zeta b(x)-\zeta c(t),~~~
   \xi^1=\zeta^2 b(x) +(1-\zeta^2) c(t)\label{x0x1.eqn}
\ee
Finally for the $\mu\nu=01$ case, assuming that $g_{01}=-\zeta$ is independent of coordinates and hence $g_{01,\lambda}=0$
we obtain, $g_{0\lambda}\xi^\lambda_{~,1}+g_{\lambda 1}\xi^\lambda_{~,0}=0$ which when expanded gives,
\be
   -(1-\zeta^2)\xi^0_{~,1}+(-\zeta)\xi^1_{~,1}+(-\zeta)\xi^{0}_{~,0}+(1)\xi^1_{~,0}=0\label{munu01.eqn}
\ee
Substituting from Eq.~\eqref{x0x1.eqn}  in this Eq.~\eqref{munu01.eqn} will imply $\partial_t c(t)=\zeta\partial_x b(x)$. The
$\mu\nu=10$ case also will give the same result. The solution of this equation is given by,
\be
   b(x)=A x +B,~~~~~~c(t)=\zeta A t +C
\ee
These solutions then give the final form for the components $(\xi^0,\xi^1)$ of the Killing vector
${\bs\xi}$ as follows,
\bearr
   &&\xi^0=A\zeta(x-\zeta t)+B\zeta-C\zeta\\
   &&\xi^1=A[\zeta^2 x+\zeta(1-\zeta^2)t]+B\zeta^2+C(1-\zeta^2)
\eearr
which depends on three parameters and therefore there are three Killing vectors.
The following choices give the Killing vectors,
\bearr
   (A,B,C)=(0,1,0)&&\rightarrow \xi_{\hat 0}\propto (1,\zeta)\label{Killt.eqn}\\
   (A,B,C)=(0,0,1)&&\rightarrow \xi_{\hat 1}\propto (-\zeta,\lambda^2)\label{Killx.eqn}\\ 
   (A,B,C)=(1,0,0)&&\rightarrow \xi_{\hat 2}\propto (x-\zeta t,\lambda^2t+\zeta x)\label{KilltL.eqn}
\eearr
Therefore the generators of symmetry operations are
\bearr
   \xi_{\hat 0}\rightarrow && \partial_t+\zeta\partial_x,\label{Killt2.eqn} \\
   \xi_{\hat 1}\rightarrow && -\zeta\partial_t+\lambda^2\partial_x, \label{Killx2.eqn}\\
   \xi_{\hat 2}\rightarrow && (x-\zeta t)\partial_t+(\lambda^2 t+\zeta x)\partial_x. \label{KilltL2.eqn}
\eearr
When the tilt parameter $\zeta$ is zero, the above generators will correspond to time translation, space translation and Lorentz transformations, 
respectively. Therefore the above generators are generators of the symmetries of the spacetime in which the dispersion of massless particles
is given by the tilted Dirac cone. Furthermore the conserve quantities in such space are not energy $H$ (corresponding to generator $\partial_t$) 
and momentum $P$ (corresponding to generators $\partial_x$). In such space the conserved quantities will be $H+\zeta P$ and $\lambda^2 P-\zeta H$ 
which of course reduce to $H$ and $P$ in the limit of $\zeta=0$. 

To construct the explicit form of the $\Lambda^x_{1+1}$, we use the generator arising from the Killing vector $\xi_{\hat 2}$ and denote it by $i\hat K$, namely
\be
   i\hat K=\left((x-\zeta t)\frac{\partial}{\partial t}+(\lambda^2 t+\zeta x)\frac{\partial}{\partial x} \right) 
\ee
which gives,
\bearr
    i\hat K x &=& \lambda^2 t+\zeta x\nn\\
    i^2\hat K^2 x &=&i \hat K(\lambda^2 t+\zeta x)=x\nn
\eearr
Repeating the above relation gives the operator equations, $i^{2n}\hat K^{2n}x=x$ and $i^{2n+1}\hat K^{2n+1}x=\lambda^2 t+\zeta x$. Therefore,
\bearr
   e^{iK\kappa}x &=&\sum_n\frac{\kappa^{2n}}{(2n)!}x+\sum_n\frac{\kappa^{2n+1}}{(2n+1)!}(\lambda^2t+\zeta x)\nn\\
   &=& \cosh\kappa x +\sinh\kappa (\lambda^2 t+\zeta x)\nn\\
   &=&\left[\cos\kappa+\zeta\sinh\kappa\right]x+\left[\lambda^2\sinh\kappa\right] t.
\eearr
Upon identification $\cosh\kappa=\gamma$ and $\sinh\kappa=-\beta\gamma$ 
this transformation of $x$ coordinate correspond precisely to the second row of Eq.~\eqref{lorentzeta.eqn}. 
Similarly by operating with $i\hat K$ on the time coordinate $t$, one obtains $i\hat Kt=x-\zeta t$ from which
it follows $i^2\hat K^2t=i\hat K(x-\zeta t)$ which is $\lambda^2t+\zeta x-\zeta(x-\zeta t)$. Using $\lambda^2+\zeta^2=1$
it finally gives $i^2\hat K^2t=t$. Therefore $i^{2n}\hat K^{2n}t=t$ and $i^{2n+1}\hat K^{2n+1}=(x-\zeta t)$. These relations 
allow us to simplify 
\be
   e^{i K\kappa}t=\left[\cosh\kappa-\zeta\sinh\kappa \right]+\left[\sinh\kappa\right] x
\ee
This will precisely correspond to the first row of Eq.~\eqref{lorentzeta.eqn}. Therefore the Killing vector $\xi_{\hat 2}$
in Eq.~\eqref{KilltL.eqn} or equivalently Eq.~\eqref{KilltL2.eqn} determines the generator of t-Lorentz transformation. 

The method of Killing to construct isometries of a given geometry (metric) is quite
general and can even be applied to space- and/or time- dependent tilt vector $\bs\zeta$.

\end{document}